\documentclass[11pt,twoside]{article}  %

\usepackage{asp2006}

\usepackage[]{graphicx}%

\begin{document}   %

\title{%
US and European Technology Roadmap for a Mid-infrared Space Interferometer
}
\author{%
{P. A.\ Schuller,\altaffilmark{{1}} }
{P. R.\ Lawson,\altaffilmark{{2}} } 
{O. P.\ Lay,\altaffilmark{{2}} } 
{A. L{\'e}ger,\altaffilmark{{1}} } and 
{S. R.\ Martin\altaffilmark{{2}} }
}
\affil{$^{1}$Institut d'Astrophysique Spatiale (IAS), 
	Universit{\'e} Paris-Sud, B{\^a}t.\ 121, 91405 Orsay Cedex, France} 
\affil{$^{2}$Jet Propulsion Laboratory (JPL), 
	MS 301-451, 4800 Oak Grove Drive, Pasadena, CA, USA}

	\begin{abstract}%
Studies of mid-infrared space interferometer concepts in the USA and in Europe have converged on a single architecture. We address the question of how the US and European communities could collaborate to advance technology efforts leading to a future space mission. 
We present the current state of the art in nulling interferometry, as demonstrated at ambient temperature and pressure in the lab, and outline required steps to demonstrate its performance under space conditions.  Design studies of a cryogenic optical test bench under vacuum have already been carried out. We highlight pre-conditions and constraints of a collaborative effort, foreseeable practical and administrative challenges, and possible strategies to meet those challenges.
\end{abstract}
\vspace{-0.40cm}


\section{State of Technology Studies}\label{sect.intro} %
JPL's Terrrestrial Planet Finder Interferometer (TPF-I) technology studies set the milestone \#3 criteria for nulling performance to achieve repeatably an interferometric null of $\le 10^{-5}$ in a spectral band width $\ge 25\%$ over 6 hours in an ambient environment. 
\emph{\textbf{
It was reached in early 2009 on the Adaptive Nuller Testbed (AdN)}} \citep{cit.peters2010pasp_adn}. %

Milestone \#4 set out to demonstrate the feasibility of four-beam nulling, the achievement of the required null stability, and the consequent detection of faint planets using approaches similar to the ones contemplated for a flight-mission. 
Experimental work was to implement this goal by the detection of a planet $\approx10^{6}$ fainter than the star at a signal-to-noise ratio of $\ge 10$ in an ambient environment. 
Phase chopping, averaging and rotation were to yield a factor $100$ in residual starlight suppression. 
Rejection levels 
were therefore expected to reach $\le 10^{-7}$
repeatably during 3 hours of operation. 
The Planet Detection Testbed (PDT) was developed to work toward this goal.  %
All the data to meet Milestone \#4 were acquired by summer 2009. 
\emph{\textbf{Milestone \#4 was formally reviewed and accepted in October 2009}} \citep{cit.martin2009jplm4r}.

Milestone \#5, planned for the year 2010, will demonstrate the effectiveness of spectral filtering to suppress instability noise. 
The expected gain in nulling depth is about one order of magnitude. 
Once this milestone is reached, further development will have to demonstrate equivalent performance under space flight conditions, i.e., \emph{\textbf{at cryogenic temperatures under vacuum}}. 

IAS has been involved in \textit{Darwin} technology studies on Achromatic Phase Shifters (APSs). 
As part of those studies, a \emph{\textbf{model for a cryogenic optical test bench under vacuum has been developed}} \citep{cit.labeque2004spie}. %

\section{Joint Efforts}\label{sect.join}
The next desired collaborative effort would be a joint study and implementation of interferometer technology and signal extraction procedures in a cryogenic system.  It would address aspects such as the following: 

\textbf{In the study phase} 
1. Review and agree on overall performance goals and conditions; 2. Develop experimental layout: light sources, detectors, 
other independent analytic tools, phase shifting device, dispersive elements, cryogenic cooling, evacuation etc.; 3. Identify sub-systems and interfaces; 4. Define interface parameters, create error budget; 5. Identify technical solutions for sub-systems. 

\textbf{In the implementation phase} 
6. Acquire, test and fully characterize sub-systems; 7. Integrate sub-systems, test and fully characterize integrated experiment; 
8. Demonstrate overall performance in compliance with goals. 

\section{Challenges and Questions}\label{sect.chall}
Given the required financial and human resources, forseeable challenges are:  1. site selection for system integration: will depend on surrounding infrastructure, availability of technical support, already available equipment etc.; 2. export/import of equipment: certain items may be rated as space and/or defense-relevant and therefore be restricted and subject to possible customs duties; 3. institutional property sovereignty: loan agreements vs.\ transfer of ownership; 4. visa regulations for visiting personnel and/or long-term exchange staff. 

Further questions include: 
A. What can be learned from earlier agreements between space agencies regarding collaboration on space missions? 
Which elements can be applied to collaborations on an institutional level at the stage of technology studies and testing? 
How and to what extent can formal agreements be made between the institutions involved? 
B. Are there (international) funding bodies which may support an overarching project as a whole and can split their funds according to local requirements of the project partners? 
C. Are there other (international) bodies which can facilitate the exchange and transfer of equipment and/or personnel by being a third partner? 

In order to present a strong case to funding agencies on both sides of the Atlantic, it is imperative to agree on common goals and build a scientific community in their support. 



\end{document}